\documentclass[12pt,letterpaper,titlepage]{article}
\usepackage[top=1in,bottom=1in,left=1in,right=1in]{geometry}
\begin{document}

\title{Probing neutron stars with gravitational waves}
\author{
Benjamin J. Owen\\
The Pennsylvania State University\\
\\
Endorsed by:\\
David H. Reitze (University of Florida)\\
Stanley E. Whitcomb (LIGO-Caltech)
}
\date{LIGO-T0900053-v1}

\maketitle

\section*{Introduction}\indent

Within the next decade gravitational-wave (GW) observations by Advanced
LIGO~\cite{LIGO} in the United States, Advanced Virgo~\cite{Virgo} and
GEO~HF~\cite{Willke:2006uw} in Europe, and possibly other ground-based
instruments will provide unprecedented opportunities to look directly into
the dense interiors of neutron stars which are opaque to all forms of
electromagnetic (EM) radiation.
The 10--$10^4$~Hz frequency band available to these ground-based
interferometers is inhabited by many neutron star mode frequencies, spin
frequencies, and inverse dynamical timescales.
GWs can provide information on bulk properties of neutron stars (masses, radii,
locations\ldots) as well as microphysics of their substance (crystalline
structure, viscosity, composition\ldots), some of which is difficult or
impossible to obtain by EM observations alone.
The former will tell us about the astrophysics of neutron stars, and the
latter will illuminate fundamental issues in nuclear and particle physics
and the physics of extremely condensed matter.
Although GW searches can be done ``blind,'' they become richer and more
informative with input from EM observations; and thus the combination of
the two is crucial for learning the most we can about neutron stars.
Healthy GW and EM observational programs must be accompanied by vigorous
theoretical research on the interface of astrophysics, gravitational
physics, nuclear and particle physics in order to extract the most from the
observations.

Advanced LIGO has been funded and it and its international partners will
not be ranked as part of the decadal review.
Our purpose here is rather to describe the neutron-star aspect of GW
science and point out observational and theory issues in EM astronomy and
astrophysics which connect to it, and thus are important to maximizing the
scientific output of the advanced detectors (LIGO, Virgo, and GEO~HF) and
the future underground GW detectors now in planning.

Opportunities to learn about neutron stars through GW and and GW/EM
observations naturally divide into four categories: supernovae, binary
mergers, starquakes, and continuous waves.
The first is addressed in another white paper~\cite{Bloom:2009vx}; here we
address the other three.

\section*{Binary Mergers}\indent

Advanced LIGO is likely to observe mergers of double neutron star (NS/NS)
binaries at a rate of a few to a few hundred per year; and
black-hole/neutron-star (BH/NS) binaries perhaps in a comparable range of
rates~\cite{Kalogera:2006uj}.
If the observed rates are at either extreme of the range, they would
reflect on extremes of compact binary populations and aspects of their
formation such as the properties of a common envelope phase and the
distribution of neutron-star birth kicks, both of which can prematurely
merge or disrupt binaries before the second neutron star or the black hole
is formed.

Individual merger signals can reveal aspects of neutron star structure:
For BH/NS mergers the tidal disruption of the neutron star can take place
at frequencies where ground-based GW detectors are most sensitive.
The latest, fully relativistic, numerical simulations of
merger~\cite{Etienne:2007jg} confirm the earlier Newtonian intuition that
the ``break frequency'' of the merger signal, the frequency at which much
of the neutron star is swallowed by the black hole and the signal becomes
very faint, is strongly correlated with the neutron star radius and
therefore can serve as a good measurement of it.
Beyond Advanced LIGO, or with a very lucky event during the time of
Advanced LIGO, a neutron star's mass can be measured to as well as 20\%
from the precessional modulation of the signal if the companion is a
rapidly rotating black hole~\cite{vanderSluys:2007st}; and a high neutron
star mass would significantly constrain the star's composition.
Recent analytical~\cite{Flanagan:2007ix} and numerical
estimates~\cite{Read:2009yp} of the effects of tidal deformations on NS/NS
orbits (and thus the phase evolution of merger signals) indicate that the
strongest signals observed by Advanced LIGO might yield measurements of
stellar radii to a precision as good as 1~km, even better than the break
frequency.
Such a radius measurement constrains to order 10\% the pressure around
twice nuclear density, a number unattainable in terrestrial laboratories
which is correlated with properties such as the incompressibility and
isospin asymmetry energy of nuclear matter.

GW and EM observations of mergers are important to each other as described
in the transient white paper~\cite{Bloom:2009vx}:
The availability of EM triggers localized in time and sky position
increases the sensitivity of GW searches, and comparison of GW and EM
observations can constrain properties of the merger and gamma-ray burst (if
any).
EM observations are important to GW mergers in another respect:
While there are several observed NS/NS systems on which to empirically base
merger rates, and eventually make population statements by comparing to
observed GW merger rates, the BH/NS event rates are estimated purely based
on theoretical population synthesis codes which contain many poorly known
factors.
If a pulsar survey, performed for instance by the Square Kilometer
Array~\cite{SKA}, reveals any BH/NS binaries it would be of great relevance
to GW merger observations.

In order to get the most science out of the observations it is also crucial
to support a program of numerical simulations of increasing complexity and
realism.
Details of the neutron star matter such as equation of state, viscosity and
other transport coefficients, photon and neutrino luminosities are modeled
at present very simply; yet they may have important signatures in the GW
emission and certainly are crucial to the EM emission.
It is also important to continue work on analytical and semi-analytical
perturbation theory, as witnessed by the recent analytical prediction of an
important tidal effect~\cite{Flanagan:2007ix} which preceded numerical
investigations~\cite{Read:2009yp} due to the computational cost of lengthy
simulations.
Advances in population simulation~\cite{Kalogera:2006uj} are also important
to extract astrophysical information from GW event rates.

\section*{Starquakes}\indent

LIGO has already placed upper limits on GW bursts associated with flares of
highly magnetized neutron stars or magnetars~\cite{Abbott:2007zzb,
Abbott:2008gj}, in some cases comparable to the $10^{45}$~erg associated
with the largest giant flare; and Advanced LIGO will be sensitive to GW
energies more than two orders of magnitude lower.
The dominant model of the flares~\cite{ThompsonDuncan} associates them with
starquakes due to sudden rearrangements of the neutron star's intense
magnetic field, quakes which would undoubtedly excite the star's
quasinormal modes and thus emit GWs.
Indeed the quasiperiodic oscillations seen in the x-ray tails of giant
flares may be modes of the crust, influenced by the magnetic
field~\cite{Israel:2005av}.
The GW upper limits are already interesting in the sense that they are just
entering the range predicted by the most extreme theoretical models.
However, it is not clear what range of GW energies would be associated with
other models of the flare, because there has been relatively little
theoretical work on the subject.
Thus there is a need for modeling to extract more meaning from current GW
observations, let alone those in the next decade.

GW searches so far have focused on frequencies tied to the quasiperiodic
oscillations seen in x-rays after the flares, on ringdowns of the
fundamental or $f$-modes which are probably the most efficient GW emitters,
and opportunistic searches for random bursts around 100~Hz where the
detectors' noise is low.
The frequency of a detected $f$-mode would strongly constrain the structure
of the star---neutron stars would be easily distinguishable from quark
stars, and the mean density can be measured to a
percent~\cite{Kokkotas:1999mn}.
The total energy radiated would reveal the mechanical efficiency of the
flare mechanism and have something to say about how much of the action
happens inside the neutron star as opposed to the magnetosphere surrounding
it.
The energy and rate at which energy is transferred to the $f$-mode from the
initial excitation could also reveal something about how the Alfv\'en
continuum of magnetic modes is coupled to the mechanical modes, and thus
how the magnetic field is distributed within the star; and any timing
offsets between GW and EM events would also constrain the possible
couplings.
Even more exciting would be a detection of a random burst not associated
with a known mode or x-ray quasiperiodic oscillation:
Such a signal could be evidence of turmoil of the magnetic field interior
to the star rather than in the magnetosphere.

The story will be similar for searches for GW bursts associated with pulsar
glitches.
Although the glitch mechanism still remains a puzzle, it can be reasonably
expected that the transfer of angular momentum to the crust from the core,
which is believed to be differentially rotating with respect to the crust,
would excite quasinormal modes and thus GW emission.
Extrapolating from~\cite{Abbott:2008gj}:
With advanced ground-based interferometers tuned to kHz frequencies, modes
excited by glitches of the nearby Vela pulsar might be observable; and
future underground detectors would be sensitive to large glitches from
pulsars within a few kpc.

For both starquake scenarios as for mergers, EM astronomy plays a crucial
role increasing the GW search sensitivity by localizing the times and sky
positions of the events.
Burst searches using non-matched filtering techniques also improve when
provided with frequency windows, such as determined by observations of
quasiperiodic oscillations or predictions of quasinormal mode frequencies.
Therefore in order to help GW observations it is important to sustain
x-ray, gamma-ray, and radio timing and rapid response capabilities.
And to extract physical information on the interiors of neutron stars it is
important to have more modeling of GW and EM emission mechanisms, and the
phenomenology of how they relate to each other.

\section*{Continuous Waves}\indent

Long-lived continuous-wave GW signals may be the hardest to detect, and
certainly they are the most computationally costly to search for; but with
$10^9$--$10^{10}$ cycles per year they could ultimately yield the most
precise information on neutron stars.
Here the connections between GW and EM astronomy are particularly tight:
Because all continuous-wave searches except monitoring of known (EM)
pulsars are severely computationally limited, electromagnetic observations
can greatly increase the sensitivities of GW searches.
The difference between the best upper limits for all-sky and known-pulsar
searches on the same set of LIGO data is nearly an order of magnitude in
strain amplitude~\cite{Abbott:2007ce, Abbott:2007tda} or a factor 100
difference in luminosity.
The all-sky survey also used much more computing power to achieve its
lesser sensitivity.

All-sky GW surveys can reveal entirely new neutron star populations.
Of the $10^8$--$10^9$ neutron stars formed in our galaxy over $10^{10}$
years, only 2000 have been identified electromagnetically---the vast
majority as pulsars---and only 20,000 could be identified as pulsars even
with the Square Kilometer Array~\cite{SKA}.
This is because most neutron stars are too old to pulse and pulsar EM
emission is relatively narrowly beamed, meaning that only a fraction of
currently EM-active pulsars can be observed.
But GW emission is beamed very little, and the emission mechanisms (which
ultimately rely on some nonaxisymmetry of the mass or momentum distribution
in the star) may not be limited in activity to the youth of the star or a
possible ``recycled by accretion'' phase as they are for pulsars.
The main selection effect is that continuous GW emission is detectable only
for spin frequencies in the ground-based detector band, about 1/5 of known
pulsars for Advanced LIGO and Advanced Virgo.
Although population studies of continuous wave sources are still in their
infancy~\cite{Knispel:2008ue}, LIGO searches are already coming within
striking distance of the most extreme scenarios~\cite{Abbott:2008rg} and
Advanced LIGO will have something to say about more likely scenarios.

Coherent integration of a year of data can allow LIGO to localize a
continuous wave source's sky location to the sub-arcsecond
level~\cite{Brady:1997ji}, facilitating followup observations to find
counterparts in x-rays, optical, and other electromagnetic wavebands.
Gravitational-wave timing of newly detected spinning neutron stars will
also aid followup searches for electromagnetic pulsations by providing spin
frequencies and frequency derivatives.
An extended observation with Advanced LIGO can also, through sphericality
of the wavefronts, obtain a distance measurement (independent of EM
observation) to about 10\%~\cite{Seto:2005gd}.

A distance measurement (whether GW-based or EM-based from a known pulsar)
allows a determination of the neutron star's quadrupolar deformation or
ellipticity, a quantity of great interest since it may shed light on
whether the neutron star is indeed made of neutrons or contains quarks or
other exotica.
Very large ellipticities are sustainable in quark models but not in normal
nuclear matter, and thus detection of a large enough ellipticity would
confirm not only the existence but also the crystalline nature of quark
matter~\cite{Owen:2005fn}.

Rapidly accreting neutron stars (whether pulsing or not) in low-mass x-ray
binaries are also very interesting targets for continuous GW searches
because the accretion is known to be asymmetric due to the occasional
observation of x-ray pulsations in some systems.
This asymmetry can lead to a GW-emitting mass asymmetry through several
mechanisms including the temperature-dependent electron capture onto nuclei
in the crust~\cite{Bildsten:1998ey}, magnetic funneling of accreted
material~\cite{Melatos:2005ez}, and sustained instability of rotational
``$r$-mode'' oscillations in the fluid below the
crust~\cite{Andersson:1998qs} (which are likely to have a briefer episode
of continuous GW emission in young neutron stars~\cite{Owen:1998xg}).
The observed lack of extremely high spin frequencies in the most rapidly
accreting neutron stars also strongly suggests that their accretion torques
are balanced by GW emission torque~\cite{Bildsten:1998ey}.
In the simplest version of this torque balance scenario, Sco~X-1 would be
detectable by advanced interferometers, while many systems would be
detectable with future underground instruments; and if the GW emission
(like the x-ray emission) also goes through active and quiescent phases,
many systems would be detectable by advanced interferometers if they go
through a GW-active phase in the next decade~\cite{Watts:2008qw}.
Determination of the neutron star spin period in Sco~X-1 or other systems
where it is completely unknown increases the sensitivity of a GW search by
a factor of a few in strain amplitude, or an order of magnitude in
luminosity.
Detection of a signal from an accreting neutron star would confirm the
hypothesis that the observed spins of older neutron stars are due to
GW/accretion torque balance, and would constrain the mechanics of accretion
as well.

Particularly interesting is the case when GW and EM observations are made
simultaneously, whether in pulsars or in the many accreting neutron stars
where the spin frequency is only roughly known (for instance through x-ray
burst oscillations):
The ratio of gravitational-wave frequency to spin frequency immediately
identifies the emission mechanism, whether a static deformation
(ellipticity) of the rotating star or an oscillating $r$-mode.
In the latter case, the frequency ratio has some dependence on the neutron
star structure~\cite{Abbott:2008fx} and thus, once observed, can be used to
measure the equation of state of dense matter.
Due to a complicated interplay of factors involving the microphysics of the
matter deep inside the star and possibly its effect on fluid and superfluid
dynamics, detection of a long-lasting $r$-mode signal from an accreting
neutron star (at any frequency) is a strong indicator for the presence of
some type of strange matter (quarks, mesons, or hyperons) in the
core (e.g.~\cite{Wagoner:2002vr, Andersson:2001ev}).
If precise EM timing is available on long timescales, even more is
possible:
A comparison between GW and EM timing can constrain the coupling between
the solid GW-emitting component of the star and whatever component of the
star and magnetosphere is emitting EM~\cite{Abbott:2008fx}; and this
comparison across a glitch may shed light on the glitch mechanism.
Future underground interferometers might also be able to observe (for the
first few days after a glitch) a continuous GW signal associated with the
fluid interior's response to the abrupt relative motion of the crust; and
this signal would carry information on the superfluidity and transport
coefficients of the matter in the interior~\cite{vanEysden:2008pd}.

What is needed to realize this potential?
Since searches for GW from known pulsars rely on coherent phase models
lasting of order a year, it is crucial to have regularly updated EM pulsar
timing data synchronous with LIGO and Virgo data runs.
With most pulsars pulsing in radio, it is important to maintain healthy
radio observatories and monitor those pulsars with high spindowns which
leave room for strong GW emission---such as PSRs J1952+3252 and J1913+1011,
which have not been observed for some time.
Also one of the most rapidly spinning down and frequently glitching
pulsars, J0537-6910, can only be timed at the moment with the Rossi X-ray
Timing Explorer satellite.
The extreme glitchiness of this pulsar indicates a substantial and quickly
changing solid component, which also indicates likely strong
gravitational-wave emission.
However it also means that without regular timing a LIGO search for J0537
would quickly become intractable, as the number of possible timings to be
used in the analysis would grow out of control.
Since RXTE is scheduled to go down soon, it is important to get a
comparable mission back up quickly.
It is also helpful to GW observations to discover as many new pulsars as
possible (in radio, x-rays, or any band), and to discover pulsations in
stars where they have not yet been observed.
Further hunts for young neutron stars and massive star forming regions will
also help GW searches by producing more targets for point or small-area
searches, which can be more sensitive than all-sky
searches~\cite{Wette:2008hg}.
Finally, it is crucial to encourage work on the theory of emission
mechanisms and all the complicated physics of neutron stars, as this is
needed to turn the combined EM and GW data streams into statements about
the properties of matter in states unreachable in terrestrial laboratories.

\bibliography{decadal.bib}

\end{document}